\documentclass[twocolumn,showpacs,preprintnumbers,amsmath,amssymb,pre]{revtex4}

\usepackage{graphicx}

\begin{document}

\preprint{}

\title{Reaction-diffusion processes on correlated and uncorrelated
scale-free networks} 

\author{Lazaros K. Gallos}
\author{Panos Argyrakis}
\affiliation{Department of Physics, University of Thessaloniki, 54124 Thessaloniki, Greece}

\date{\today}

\begin{abstract}
We compare reaction-diffusion processes of the $A+A\to 0$ type on scale-free networks
created with either the configuration model or the uncorrelated configuration model.
We show via simulations that except for the difference in the behavior of the two models, different
results are observed within the same model when the minimum number of connections
for a node varies from $k_{\rm min}=1$ to $k_{\rm min}=2$. This difference is attributed to
the varying local properties of the two systems. In all cases we are able to
identify a power law behavior of the density decay with time with an exponent $f>1$,
considerably larger than its lattice counterpart.
\end{abstract}


\maketitle

Recently, a growing interest for dynamic processes taking place on scale-free networks
has arisen \cite{AB,DM}. A scale-free network is one where the node connectivity $k$ (number of links
on a node) follows a power-law distribution of the form
\begin{equation}
\label{Pk}
P(k) \sim k^{-\gamma} \,,
\end{equation}
where $\gamma$ is a positive number, typically in the range $2<\gamma<4$.
In this frame, we recently presented \cite{GA} simulation results for reaction-diffusion processes both of the type
$A+A\to 0$ and $A+B\to 0$, where the substrate for diffusion is a scale-free network.
Following this, Catanzaro et al. \cite{CBPSa} developed an elegant theory for the $A+A\to 0$
process, which applies on networks created with the uncorrelated configuration model (UCM) \cite{CBPSb}.
Their analytic results were found to be in good agreement to their simulations, but were
deviating from the results in Ref. \cite{GA}, where the networks were created with
the configuration model (CM). The authors attributed the difference of the results on
the different method of preparing the networks. In this Brief Report we directly compare
results for both the UCM and CM models.

\begin{figure}
\includegraphics{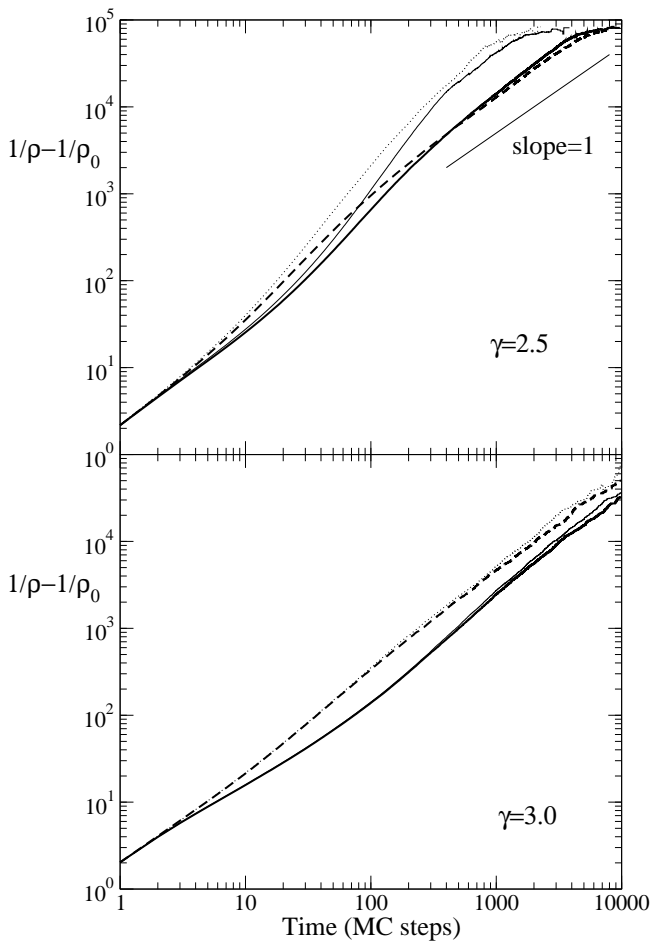}
\caption{\label{fig1} Plot of the reaction progress $1/\rho-1/\rho_0$, as a function of
time for the A+A$\to$0 reaction on scale-free networks of (a) $\gamma=2.5$, and (b) $\gamma=3.0$.
Results correspond to CM (thin lines) and UCM (thick lines) models, with $k_{\rm min}=1$ (solid lines)
and $k_{\rm min}=2$ (dashed lines).
The initial density was $\rho_0=0.5$. All results correspond to networks of $N=10^6$ nodes.
}
\end{figure}

The Configuration Model (CM) has become more or less the standard for simulating networks of a given $\gamma$
value in the literature. For each one of the $N$ system nodes a random $k$ value is assigned, drawn from
the probability distribution function of Eq.~(\ref{Pk}). Pairs of links are then randomly chosen between the nodes,
taking care that no double links between two nodes or self-links in the same node are established. When the
maximum value of $k$ is not pre-defined, then the natural upper cutoff scales with the system size as
$k_{\rm max}\sim N^{1/(\gamma-1)}$. This model is known to create correlations in the connectivity distribution
between the system nodes for $\gamma<3$, in the sense that the average connectivity of a node's neighbors depends on its
rank $k$. In other words, nodes that are highly connected prefer to attach to nodes with lower $k$-values, rather than to
equally well-connected nodes.
The Uncorrelated Configuration Model (UCM) uses the same construction algorithm, with the difference
that the upper cutoff is fixed in advance to $k_{\rm max}=N^{1/2}$. Then, it has been shown \cite{CBPSb} that connectivity
correlations dissapear and the average connectivity for the neighbors of any node is constant.

In Ref.~\cite{GA} we had found that the reaction rate was evolving surprisingly faster than
on regular lattices. The variation of the particle concentration $\rho(t)$ was still following a
power law with time $t$ of the form
\begin{equation}
\label{rho}
\frac{1}{\rho(t)}-\frac{1}{\rho_0} \sim t^f \,,
\end{equation}
but with a value $f>1$ for $\gamma<3.5$. This result was valid asymptotically within the simulation accuracy, and before
finite-size effects settled in. The
networks that we had used had been created with the CM model with a minimum value for the connectivity
of a node (lower cutoff) $k_{\rm min}=1$. Since this process may create isolated clusters, we only had
used the largest cluster which, depending on the value of $\gamma$, would span from 35\% to 100\% of the
system nodes.

In Ref.~\cite{CBPSa} the $A+A\to 0$ process was studied, and it was analytically found that for an infinite size
network the exponent $f$ of Eq.~(\ref{rho}) is given by $f=1/(\gamma-2)$ when $2<\gamma<3$, i.e.
again $f>1$. For $\gamma=3.0$ the behavior was predicted analytically to be $1/\rho\sim t\ln{t}$.
However, for finite size networks and long times the behavior of $1/\rho$ is masked for all $\gamma$ by the mean-field
exponent $f=1$, which seems to also be valid asymptotically from the simulation results. Both the
analytic solution and the simulations in that paper were based on networks created under the UCM
model, with a lower cutoff value of $k_{\rm min}=2$. The authors of Ref.~\cite{CBPSa} reported that
in their simulations it was not possible to find a noticeable regime with $f>1$, and they attribute the
discrepancy between the two studies solely in the different network creation method. We here argue that a
power law regime with $f>1$ is indeed present and clearly identified in all cases. Additionally, a very
important factor is the lower cutoff value, $k_{\rm min}$.

In the present Report we present and compare simulation results for both the CM and UCM models with $k_{\rm min}=1$
and $k_{\rm min}=2$. Results for the time evolution of the particle density in all four possible cases
are shown in Fig.~\ref{fig1}.
All four curves in the case of $\gamma=2.5$ (Fig.~\ref{fig1}a) behave differently from each other. The CM and UCM models
clearly yield different results, but even within UCM or CM the curves for $k_{\rm min}=1$ and $k_{\rm min}=2$ deviate
from each other. 
The main difference is that the crossover to the power law behavior for $k_{\rm min}=1$ appears later in time,
and the asymptotic mean-field behavior ($f=1$) also exhibits itself roughly one decade later.

The case of $\gamma=3.0$ (Fig.~\ref{fig1}b) is simpler. For this $\gamma$
value the two models (CM and UCM) are expected to coincide, since in general the natural upper cutoff in 
the CM model scales as $k_c\sim N^{1/\gamma-1}$, and for $\gamma=3.0$ the value $k_c=N^{1/2}$ is exactly the
same as in the UCM model. This coincidence is shown to be valid in the figure. However, the results for $k_{\rm min}=1$
are still different than the results for $k_{\rm min}=2$. The crossover to the power-law behavior is not as prominent as in
the case of $k_{\rm min}=1$. 

\begin{figure}
\includegraphics{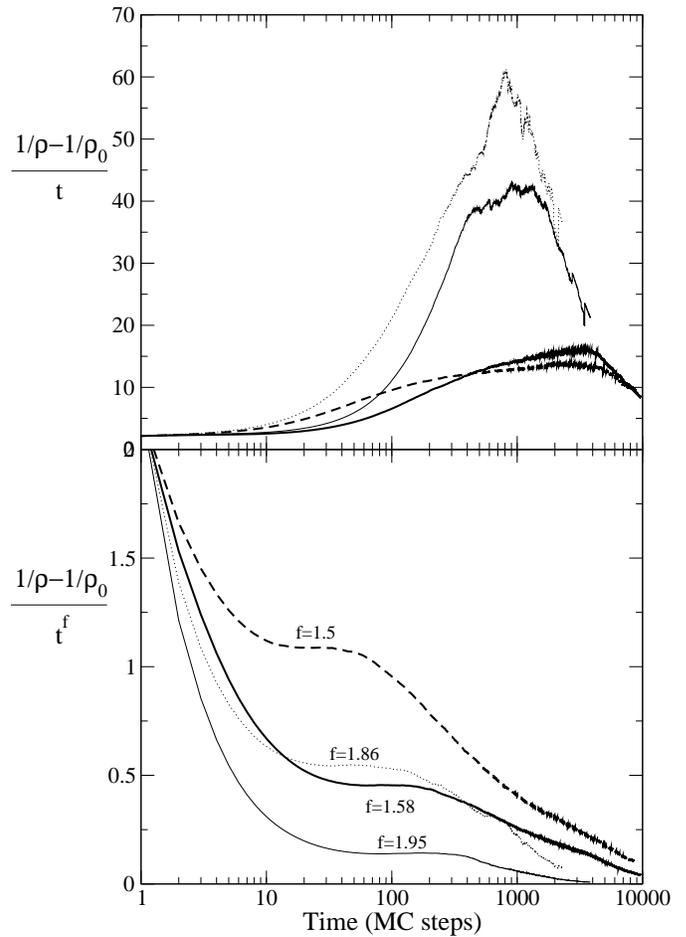}
\caption{\label{fig2} Results for networks with $\gamma=2.5$.
Plots of inverse particles density ($1/\rho-1/\rho_0$)
divided by (a) $t$ and (b) $t^f$, as a function of time. Thick lines correspond to the UCM and thin lines to the CM model
while in solid curves $k_{\rm min}=1$ and in dashed lines $k_{\rm min}=2$. In (b) the $f$ values
are shown on the plot.}
\end{figure}

We then assess whether there exists an observable power-law behavior in these curves and whether asymptotically
a linear regime masks this power-law. The results for $\gamma=2.5$ are presented in Fig.~\ref{fig2}. In the first
plot we divide the curves of Fig.~\ref{fig1}a by the linear function $t$. If asymptotically the expected behavior
is linear ($f=1$ in Eq.~\ref{rho}) then we expect that this division will yield a constant value, i.e. a
horizontal line parallel to the x-axis. The only case which is close to this behavior is that of the UCM model
and $k_{\rm min}=2$ (thick dashed line in the plot). Even then, the result is not entirely satisfactory, and the
curve seems to have a slope greater than 0. Notice also that although from Fig.~\ref{fig1}a one can visually get the
impression that a slope of 1 is describing quite accurately this curve, the more detailed analysis in Fig.~\ref{fig2}a reveals
that this is not true. In all other
cases, there are only weak hints for a parallel line, but it cannot be argued that this linear relation is in
general valid, and this is why it was not observed in Ref.~\cite{GA}.
The abruptly falling curves at longer times are due to very low particle densities, where
less than 10 particles remain in the system.

\begin{figure}
\includegraphics{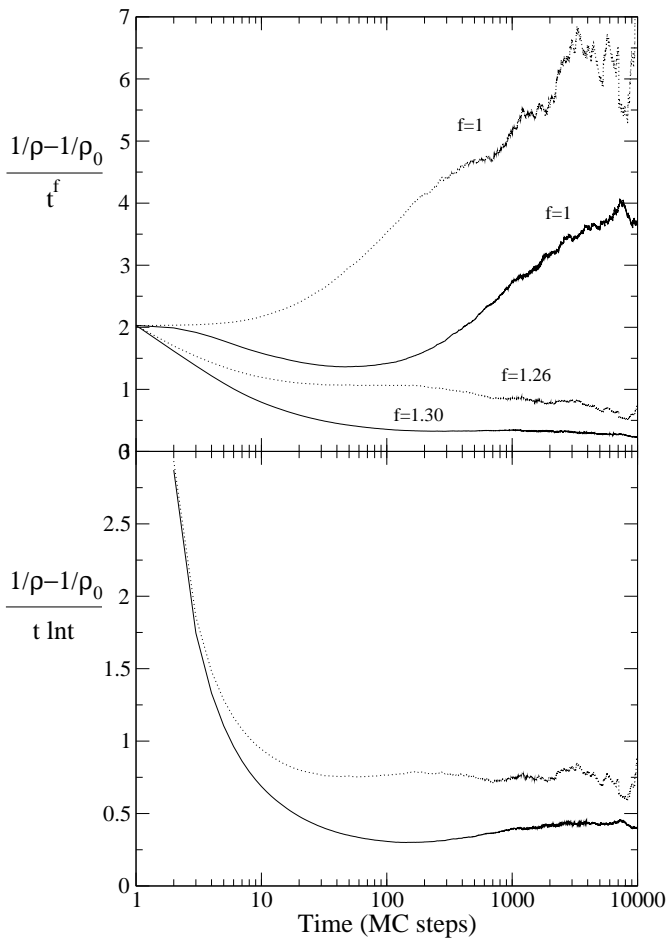}
\caption{\label{fig3} Results for networks with $\gamma=3.0$. Plots of inverse particle density ($1/\rho-1/\rho_0$)
divided by (a) $t^f$ and (b) $t \ln{t}$, as a function of time. The CM and UCM models yield the same curves, so we
only present results for CM with $k_{\rm min}=1$ (solid lines) and $k_{\rm min}=2$ (dashed lines).}
\end{figure}

In Fig.~\ref{fig2}b, we also test the hypothesis for power-law behavior by dividing the raw data
of Fig.~\ref{fig1}a by $t^f$. Since curves corresponding to different models exhibit varying exponents
$f$ we choose the value of $f$ that maximizes the extent of the horizontal part in each line.

The presence of a power law regime is clear in all occasions, but it is also noteworthy that the validity of the power law
lies in a narrower time range for $k_{\rm min}=2$, and especially from the UCM model, rendering this estimation more
difficult. Most probably, this is the reason that the authors of Ref.~\cite{CBPSa} did not characterize this regime.
The case where in a simulation a particular power law shows only in a very limited time domain is not unusual,
see e.g. the Zeldovich behavior of the $A+B$ reaction on a 3-dimensional lattice,
where the density exponent has
a nominal value of $f=0.75$, which shows only in one time decade \cite{Argyrakis}.
The slopes of the curves for $k_{\rm min}=1$ are
consistently larger compared to the slopes of the $k_{\rm min}=2$ curves. We also note that the
analytical prediction of Ref.~\cite{CBPSa} predicts a slope of $f=2.0$, which is very close to the slope of
the CM model for $k_{\rm min}=1$, but deviates remarkably from the slope of the UCM model, although this
prediction was based on the uncorrelated network hypothesis.

We have already seen that when $\gamma=3.0$ the CM and UCM models are identical. For clarity in Fig.~\ref{fig3}
we analyze only the results of CM. Again, the hypothesis that asymptotically there is a linear increase is not
directly supported from our simulation results. On the contrary, if we consider a power law fit for the intermediate time regime,
we observe an extended horizontal part that spans a few decades. The slope for both $k_{\rm min}=1$ and 2
are similar and close to $1.3$.
An increase
in the network size $N$ has been shown to further extend the region of validity for the power-law behavior in the $f>1$ regime (see e.g. Fig.~1 of \cite{GA}
and Fig.~6 of \cite{CBPSa}).

The analytic considerations in Ref.~\cite{CBPSa} for $\gamma=3.0$ predict a logarithmic correction
($1/\rho\sim t\ln{t}$). In general, it is not easy to distinguish between this form and a power law curve.
However, when we divide our raw simulation data with the $t\ln{t}$ function in Fig.~\ref{fig3}b, we observe
that for $k_{\rm min}=2$ there is an excellent agreement, but this function fails to reliably describe the
results for $k_{\rm min}=1$.

The most probable explanation for the difference in the results for the varying $k_{\rm min}$ values is that
the local environment in the case of $k_{\rm min}=1$ is remarkably different than when $k_{\rm min}=2$.
Although $k_{\rm min}=2$ is known to yield one giant connected cluster \cite{Cohen}, its structural
characteristics seem to be modified, basically because of the tree-like features (many nodes are connected to
just one neighbor). Bottlenecks and revisitations are, thus, more frequent than in the case where a larger lower cutoff is used.

In short, the value of the exponent characterizing a reaction-diffusion process on scale-free
networks depends not only on the network construction model, but is also sensitive to the lower cutoff value
for the connectivity $k_{\rm min}$, although such a dependence has not been predicted analytically.
More importantly, we have verified the existence of a power law regime that is large enough
to be observable in all cases, while a careful analysis also revealed that an asymptotic linear behavior is not
in general valid.

\end{document}